# All-optical frequency resolved optical gating for isolated attosecond pulse reconstruction

Zhen Yang[1], Wei Cao[1,*], Xi Chen[1], Jie Zhang[1], Yunlong Mo[1], Huiyao Xu[1], Kang Mi[1], Qingbin Zhang[1], Pengfei Lan[1], and Peixiang Lu[1,2]

[1]Wuhan National Laboratory for Optoelectronics and School of Physics, Huazhong University of Science and Technology, Wuhan, 430074, china
[2]Hubei Key Laboratory of Optical Information and Pattern Recognition, Wuhan Institute of Technology, Wuhan 430205, China
*Corresponding author: weicao@hust.edu.cn, lupeixiang@hust.edu.cn

**We demonstrate an all-optical approach for precise characterization of attosecond extreme ultraviolet pulses. Isolated attosecond pulse is produced from high order harmonics using intense driving pulse with proper gating technique. When a weak field is synchronized with the driver, it perturbs the harmonics generation process via altering the accumulated phase of the electron trajectories. The perturbed harmonic spectrum can be formulated as a convolution of the unperturbed dipole and a phase gate, implying the validity of complete reconstruction of isolated attosecond pulses using conventional frequency resolved optical gating method. This *in situ* measurement avoids the central momentum approximation assumed in the widely used attosecond streaking measurement, providing a simple and reliable metrology for isolated attosecond pulse.**

Utilizing the process of the high harmonic generation (HHG) for producing single attosecond ($10^{-18}$ as) pulses [1-3] opens the gate of observing ultrafast dynamics in atoms, molecules and solids [4-11]. Naturally, it is essential to acquire the complete information of the advanced metrological tool before application. However, for such ultrashort pulses it is still challenging to directly measure its temporal structure, as there are neither means with sufficient temporal resolution for attosecond pulse sampling nor available nonlinear crystals for autocorrelation measurement. Fortunately, the short temporal scale of the pulse always imply broader frequency distribution based on the Fourier transform properties, which brings great convenience to characterize attosecond pulses in frequency domain.

In the past two decades, several schemes have been developed for attosecond pulses characterization in frequency domain. They are generally divided into two categories: *ex situ* and *in situ* measurements [12]. For the former one the generation and diagnosis of attosecond pulses are performed in different locations, while for the later one both the generation and measurement are carried out in the same target cell. The *ex situ* scheme, such as reconstruction of attosecond beating by interference of two-photon transitions (RABBITT) [13], attosecond streak camera [14] and single-color two-photon above-threshold ionization [15], converts extreme ultraviolet (EUV) pulses information into photoelectron momentum and requires precise measurement of the photoelectron spectrum using sophisticated charged particle detecting systems. Nowadays, the most used method for complete characterization of attosecond pulse is the FROG-CRAB method [16]. This method is valid as long as the central momentum approximation is applicable [16]. Nevertheless, for attosecond pulses with low photon energy near the ionization threshold of matters or ultra-broad spectral bandwidth, its application is limited due to the breakdown of the central momentum approximation. As for the *in situ* scheme, it has previously been demonstrated by means of introducing an external second-harmonic pulse for breaking the symmetry, and the emission time of HHG is encoded in the even-order harmonics [17]. Later, Kim *et al* demonstrate an all-optical scenario for characterizing attosecond pulses in space-time domain [18]. In that scheme, a weak perturbing pulse propagates non-collinearly with the fundamental with a small deflection angle, a FROG like measurement is performed in the space-momentum domain, allowing one to measure the accurate spatial phase of EUV pulse in the far field. The relative spectral phase in this scheme is acquired via the delay dependent oscillation at different photon energy. Both *in situ* methods only measure the group delay in HHG, and more complex phase information such as high order dispersion in attosecond pulse calls for more accurate phase retrieval method.

In the present work, we demonstrate an all-optical approach for accurate characterization of the temporal structure of isolated attosecond pulses. Our approach is inspired by the conventional frequency-resolved optical gating (FROG) [19], which has been wildly applied to femtosecond pulses characterization. The conventional FROG trace can be written in the following form:

$$S_0(\omega,\tau) = \left|\int_{-\infty}^{+\infty} E(t)G(t,\tau)e^{i\omega t}dt\right|^2 \quad (1)$$
$$= \left|\tilde{E}(\omega) \otimes \tilde{G}(\omega,\tau)\right|^2$$

where $\tau$ represents the relative delay between the unknown field $E(t)$ and the gate $G(t,\tau)$ operating on the field to be characterized. The trace can be also expressed in frequency domain as shown in the second row of Equation (1), where $\otimes$ is the convolution operator. $\tilde{E}(\omega)$ and $\tilde{G}(\omega,\tau)$ are the Fourier transform of $E(t)$ and $G(t,\tau)$, respectively. Consequently, waveforms of both $E(t)$ and $G(t,\tau)$ can be extracted via the efficient principal components generalized projection algorithm (PCGPA) [20]. The similar reconstruction procedure is also applied in attosecond pulse characterization [15,16], where the photoelectrons is liberated by EUV pulses form the FROG trace.

In our current all optical scheme, a strong fundamental field is used to drive high harmonic radiation [21] and isolated attosecond pulse can be generated if an optical gating is property created [22-29]. An additional low-intensity (<10$^{-2}$) field collinearly propagates with the driving field and perturbs the dynamics of free electrons, which alters the accumulated phase of free electron acquired in the driving field. By scanning the relative delay between the perturbing and driving fields, the spectrum of the attosecond pulse shows strong modulation in both amplitude and central photon energy with respect to delay (see Fig. 1(a)).

The time-dependent dipole moment after introducing the sufficiently weak perturbing field can be written as [16]:

$$d(t,\tau) = \sum_j d_0^j(t)e^{-i\sigma^j(t,\tau)} + c.c. \quad (2)$$

$d_0^j(t)$ stands for the dipole moment with only driving field present, the sup-script j stands for the corresponding quantum trajectory. $\sigma^j(t,\tau) = -\int_{t'}^{t(t')}[P_S - A_0(t'')]A_S(t'',\tau)dt''$ is the additional phase induced by the perturbing pulse with t and t' representing the ionization and recombination time, respectively. $P_S$, $A_0$ and $A_S$ correspond to the canonical momentum, the vector potential of the driving field and perturbing field. If a single attosecond pulse is generated, i.e. a single quantum trajectory is considered, equation (2) is reduced to: $d(t,\tau) = d_0^1(t)e^{-i\sigma^1(t,\tau)} + c.c.$. The two-dimension spectrogram trace is obtained by performing the Fourier transform of the dipole acceleration:

$$S(\omega,\tau) = \omega^4 \left|\tilde{d}_0^1(\omega) \otimes H(\omega,\tau)\right|^2 \quad (3)$$

where $H(\omega,\tau) = F\left[e^{-i\sigma^1(t,\tau)}\right]$ is the Fourier transform of the additional phase induced by the perturbing field. Equation (3) is essentially identical to Equation (1) with the phase term $e^{-i\sigma^1(t,\tau)}$ serving as a gate in a FROG trace. This implies that both $d_0^1(t)$ and $e^{-i\sigma^1(t,\tau)}$ can be extract from the spectrogram trace using the standard FROG algorithm.

Figure 1 shows the principle of the all-optical FROG method for attosecond pulse reconstruction. When a few cycle laser pulse drives HHG, the cut-off harmonic shows a continuous structure, supporting the generation of isolated attosecond pulse (trace B in Fig. 1(a)). If a weak field is introduced the high harmonic spectrum is perturbed and sensitively depends on the relative delay between the two pulses. We performed the simulation using Lewenstein's model [1]. Here we add a window function on the radiation dipole moment of electron for selecting a single attosecond pulse in short trajectory region, i.e., mimic the isolate attosecond pulse generation. The calculated two-dimension neon HHG spectrum generated by $2\times10^{14}$ W/cm$^2$, 5 fs, 750nm driving pulse and $2\times10^{11}$W/cm$^2$, 5 fs, 750nm perturbing pulse is shown in Fig 1.(b). Comparing the trace in Fig 1(b) with a typical FROG-CRAB trace, the major difference between them is that the FROG-CRAB trace shows a clear delay dependent central momentum shift, while the strong yield modulation is a more prominent effect in the trace of the current method. Assuming the spectrogram in Fig 1(b) can be expressed by equation (3), we can then retrieve the attosecond pulse using the PCGP algorithm. The retrieved results are shown in Fig. 1(c). It shows that both the intensity profile and phase of the retrieved attosecond pulse show good agreement with that of the original one, proving the validity of our method.

Our experimental set-up is based on a Mach-Zehnder interferometer as shown in Fig. 2(a). A few cycle near infrared pulse centered at 750nm is split by a 20% reflection beam splitter. The transmission beam serves as a driving field, which passes through the double optical gating (DOG) optics [28, 29] for single attosecond pulse generation. A delay stage and an iris are settled in the reflection beam path for adjusting the relative delay between two pulses and the intensity of perturbing field, respectively. After that, two beams are recombined by another 20% reflection beam splitter and focused into argon gas cell for high harmonic generation (HHG). In the gas cell, the effective intensity in the DOG gate is estimated to be $2\times10^{14}$ W/cm$^2$, while the leak intensity of the perturbing field is less than 1%. The residual femtosecond pulses after HHG cell are blocked by a 200nm thick aluminum film. Finally, the spectrum of the attosecond pulse is detected by an EUV spectrometer.

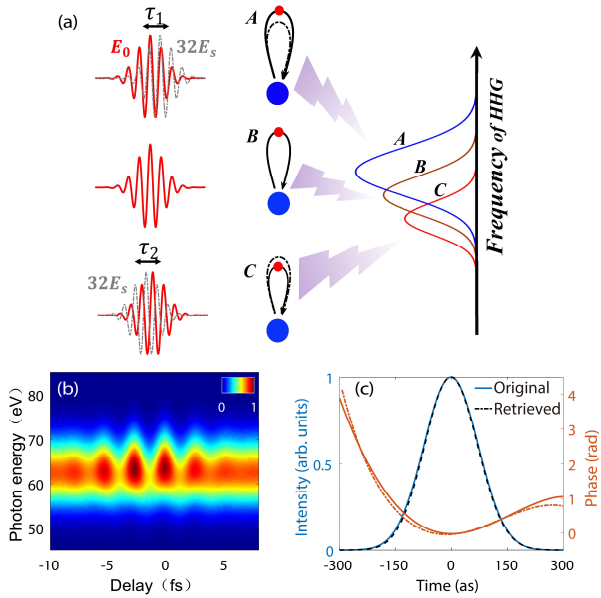

Fig. 1. Principle of the all-optical FROG measurement of isolated attosecond pulse. (a) When a weak pulse $E_s$ is introduced to perturb the high harmonic generation process (trace B) driven by the strong pulse $E_0$, both the amplitude and frequency of the radiated attosecond pulse will be altered depending on the relative delay. (b) the calculated two-dimension HHG spectrum. (c) the reconstructed (solid lines) and original (dashed lines) attosecond pulse show good agreement in both the intensity profile and temporal phase.

When the DOG optics are properly tuned, supercontinuum harmonic radiation is generated when only the intense driving laser beam is present as shown in the inset in Fig. 2(a). An attosecond streak experiment indicates that a single attosecond pulse has been successfully generated (see Fig. 3). As a delayed weak perturbing field is introduced, both central photon energy and amplitude of the HHG spectrum are oscillating prominently as a function of delay as shown in Fig. 2(b). Figure 2(b) resembles the main features of a perturbed harmonic spectrogram as predicted by simulation in Fig. 1(b). Following the procedure introduced before, we can reconstruct the temporal structure of the dipole moment, i.e. the attosecond pulse. Figure 2(d) shows the retrieved temporal structure of a single attosecond pule with full width at half maximum (FWHM) of 272 as.

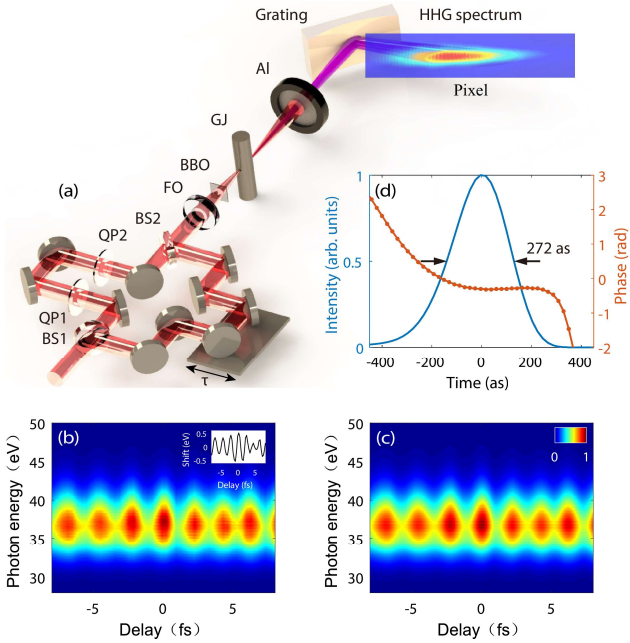

Fig. 2 (a) Experimental layout of the all-optical FROG for isolated attosecond pulse reconstruction. BS: beam splitter, QP1: 330 um quartz plate, QP2: 470 um quartz plate, FO: focusing optics, BBO: barium borate crystal, GJ: gas jet, Al: aluminum film. (b) Measured spectrogram with a delay step size of 400 as. The inset shows the oscillating central photon energy. (c) Retrieved trace. (d) The retrieved temporal profile (blue solid line) and its phase (orange dotted line) of the attosecond pulse.

We also perform an attosecond streaking measurement to compare with our all optical FROG method. The experimental set-up is similar to that in [30]. The single attosecond pulse and a delayed streaking field is focused into an effusive jet backed with neon gas, and a field free time of flight spectrometer is used to measure the photoelectron spectrum. Figure 3(a) and 3(b) shows the measured and reconstructed streaking trace, respectively. The temporal domain information of the retrieved isolated attosecond pulse is shown in Fig. 3(c). The intensity profile and phase of the retrieved attosecond pulse from the two methods gives rather similar results. It should be noted that the central momentum approximation used in the FROG-CRAB measurement may bring error in retrieving attosecond pulses. In contrast, the all-optical FROG method does not require this approximation and therefore is an accurate technique without any critical defect in reconstruction process except requiring the perturbing field weak enough.

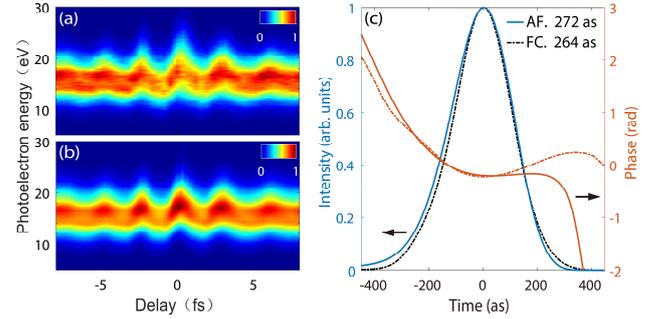

Fig. 3 FROG-CRAB measurement of the isolated attosecond pulse. (a) Measured FROG-CRAB trace of neon gas. (b) Retrieved FROG-CRAB trace. (c) The temporal information of attosecond pulse retrieved via all-optical FROG measurement (solid lines) and FROG-CRAB measurement (dashed lines). In both measurements, the attosecond pulse is generated under the same experimental conditions as Fig. 2.

In general, any gating technique should be applicable for the current all-optical method. In Fig. 1 we choose the amplitude gating technique for convenience to illustrate the principal of our method. To verify the validity of our experimental results, we also performed the numerical simulation using double optical gating with parameters similar to that in the experiment. The time-dependent ellipticity of the driver is formed by two counter rotating 5 fs, 750nm few cycle pulses, and the linear polarized gate for isolated pulse generation has an intensity of $2.35\times10^{14}$W/cm$^2$. The perturbing pulse has an intensity of $5\times10^{11}$W/cm$^2$. The simulated trace is shown in Fig. 4(a). It exhibits a rather similar trace to the experimental one in Fig. 2(b). Figure 4(c) shows the retrieved attosecond pulse, which is consistent with the original one, confirming that the current method is applicable for DOG technique. Equation (3) indicates that our scheme is applicable if a single trajectory is selected. To test the generality of our method, we consider the case where both short and long trajectories are existing (this can be realized by adding two window functions in the radiation dipole). The simulated perturbed high harmonic spectrum is shown in Fig. 4(b). The interference between the two trajectories leads to modulation along the energy axis. Since more than one trajectories are included, strictly speaking equation (3) does not holds and the retrieved time profile of the attosecond pulses deviates from that of the original one slightly (see Fig. 4(d)). Nevertheless, the phase information of the two pulses has been properly recovered and the dual pulse structure is revealed qualitatively, which is indeed valuable information for attosecond experiments.

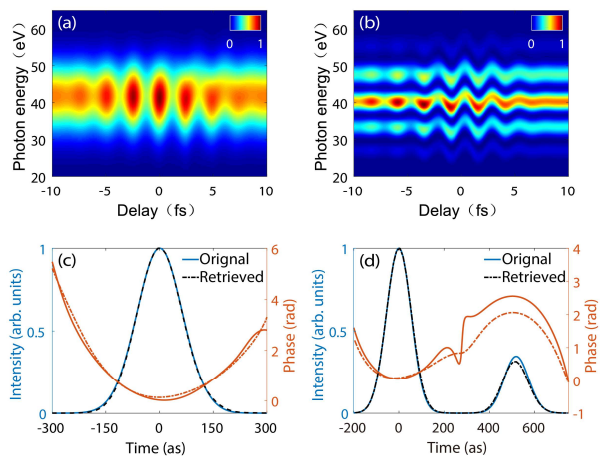

Fig. 4 Numerical analysis of all-optical FROG measurement of attosecond pulses generated from argon using DOG. (a) Calculated two-dimension harmonic spectrum when a single trajectory is selected with a window function. (c) The retrieved (solid lines) isolated attosecond pulse from the trace (a) as compared with the original (dashed lines) one. (b) Calculated two-dimension harmonic spectrum when both short and long trajectories are selected. (d) The retrieved (solid lines) attosecond double pulse from trace (b) as compared with the original one (dashed lines).

In conclusion, an all-optical FROG like measurement has been successfully demonstrated for complete reconstruction of isolated attosecond pulse. It utilizes a weak field to perturb a single HHG trajectory and forms a FROG trace in the EUV domain. The current approach does not assume the central momentum approximation generally used in FROG-CRAB measurement, and thus can be potentially used to diagnose attosecond pulses with more flexible spectral range. It is a convenient and reliable all-optical technique for isolated attosecond pulse characterization with broad application prospects.

**Funding.** National Natural Science Foundation of China (11774111, 91950202) and National Key Research and Development Program(2017YFE0116600), the Fundamental Research Funds for the Central Universities(2017KFYXJJ142) and International Cooperation program of Hubei Innovation Fund (2019AHB052).

**Disclosures**. The authors declare no conflicts of interest.

## References

1. M. Lewenstein, Ph. Balcou, M. Yu. Ivanov, A. L'Huillier, and P. B. Corkum, Phys. Rev. A **49**, 2117 (1994).
2. F. Krausz, and M. Y. Ivanov, Rev. Mod. Phys. **81,** 163–234 (2009).
3. P. Lan, P. Lu, W. Cao, Y. Li, and X. Wang, Phys. Rev. A, **76**, 011402(2007).
4. M. Hentschel, R. Kienberger, Ch. Spielmann, G. A. Reider, N. Milosevic, T. Brabec, P. Corkum, U. Heinzmann, M. Drescher, and F. Krausz, Nature **414**, 509–513 (2001).
5. R. Kienberger, E. Goulielmakis, M. Uiberacker, A. Baltuska, V. Yakovlev, F. Bammer, A. Scrinzi, T. Westerwalbesloh, U. Kleineberg, U. Heinzmann, and M. Drescher, Nature **427**, 817 (2004).
6. E. Goulielmakis, Z. Loh, A. Wirth, R. Santra, N. Rohringer, V. S. Yakovlev, S. Zherebtsov, T. Pfeifer, A. M. Azzeer, M. F. Kling, S. R. Leone, and F. Krausz, Nature **466**, 739–743 (2010).
7. H. J. Wörner, J. B. Bertrand, D. V. Kartashov, P. B. Corkum, and D. M. Villeneuve, Nature **466**, 604–607 (2010).
8. G. Sansone, F. Kelkensberg, J. F. Pérez-Torres, F. Morales, M. F. Kling, W. Siu, O. Ghafur, P. Johnsson, M. Swoboda, E. Benedetti, and F. Ferrari, Nature **465**, 763 (2010).
9. M. Schultze, K. Ramasesha, C. D. Pemmaraju, S. A. Sato, D. Whitmore, A. Gandman, J. S. Prell, L. J. Borja, D. Prendergast, K. Yabana, and D. M. Neumark, Science **346**, 1348 (2014).
10. M. Garg, M. Zhan, T. T. Luu, H. Lakhotia, T. Klostermann, A. Guggenmos, and E. Goulielmakis, Nature **538**, 359–363 (2016).
11. Z. Tao, C. Chen, T. Szilvási, M. Keller, M. Mavrikakis, H. Kapteyn, and M. Murnane, Science **353**, 62–67 (2016).
12. K. T. Kim, D. M. Villeneuve, and P. B. Corkum, Nature Photon. **8**, 187–194 (2014).
13. P. M. Paul, E. S. Toma, P. Breger, G. Mullot, F. Augé, Ph. Balcou, H. G. Muller, and P. Agostini, Science **292**, 1689–1692 (2001).
14. J. Itatani, F. Quéré, G. L. Yudin, M. Yu. Ivanov, F. Krausz, and P. B. Corkum, Phys. Rev. Lett. **88**, 173903 (2002).
15. A. Kosuge, T. Sekikawa, X. Zhou, S. Adachi, and S. Watanabe, Phys. Rev. Lett. **97**, 263901 (2006).
16. Y. Mairesse, and F. Quéré, Phys. Rev. A **71**, 011401(R) (2005).
17. N. Dudovich, O. Smirnova, J. Levesque, Y. Mairesse, M. Yu. Ivanov, D. M. Villeneuve, and P. B. Corkum, Nature Phys. **2**, 781–786 (2006).
18. K. T. Kim, C. Zhang, A. D. Shiner, S. E. Kirkwood, E. F., G. Gariepy, A. Naumov, D. M. Villeneuve, and P. B. Corkum, Nature Phys. **9**, 159–163 (2013).
19. D. J. Kane, and R. Trebino, IEEE J. Quantum Electron. **29**, 571–579 (1993).
20. D. Kane, IEEE J. Quantum Electron. **35**, 421 (1999).
21. Corkum, P. B., Phys. Rev. Lett. **71**, 1994–1997 (1993).
22. Ch. Michael, K. Zhao, and Z. Chang, Nature Photonics **8**, 178 (2014).
23. E. Goulielmakis, M. Schultze, M. Hofstetter, V. S. Yakovlev, J. Gagnon, M. Uiberacker, A. L. Aquila, E. M. Gullikson, D. T. Attwood, R. Kienberger, F. Krausz, and U. Kleineberg, Science **320**, 1614–1617 (2008).
24. P. B. Corkum, N. H. Burnett, and M. Y. Ivanov, Opt. Lett. **19**, 1870 (1994).
25. I. P. Christov, M. M. Murnane, and H. Kapteyn, Phys. Rev. Lett. **78**, 1251 (1997).
26. P Lan, P Lu, W Cao, Y Li, and X Wang, Phys. Rev. A **76**, 051801 (2007).
27. W Cao, P Lu, P Lan, X Wang, and G Yang, Phys. Rev. A **74**, 063821 (2006).
28. Z. Chang, Phys. Rev. A **76**, 051403(R) (2007).
29. Q. Zhang, P. Lu, P. Lan, W. Hong, and Z. Yang, Optics Express **16**, 9795 (2008).
30. X. Feng, S. Gilbertson, H. Mashiko, H. Wang, S. D. Khan, M. Chini, Y. Wu, K. Zhao, and Z. Chang, Phys. Rev. Lett. **103**, 183901 (2009).